\documentclass{vgtc}     

\ifpdf%
  \pdfoutput=1\relax                   %
  \usepackage{graphicx}                %
  \DeclareGraphicsExtensions{.pdf,.png,.jpg,.jpeg} %
\else%
  \usepackage{graphicx}                %
  \DeclareGraphicsExtensions{.eps}     %
\fi%

\graphicspath{{figures/}{pictures/}{images/}{./}} %

\usepackage{microtype}                 %
\PassOptionsToPackage{warn}{textcomp}  %
\usepackage{textcomp}                  %
\usepackage{mathptmx}                  %
\usepackage{times}                     %
\usepackage{cite}                      %
\usepackage{tabu}                      %
\usepackage{booktabs}                  %

\onlineid{0}

\vgtccategory{Research}

\vgtcinsertpkg

\usepackage{amsmath}
\usepackage{graphicx}
\usepackage{balance}  %
\usepackage{adjustbox}
\usepackage{footmisc}
\usepackage{soul}
\usepackage{hyperref}
\usepackage{enumitem}
\usepackage{multirow}
\usepackage{listings}
\usepackage{cleveref}
\usepackage{array}
\usepackage{tabularx}
\usepackage{booktabs} %
\usepackage{xspace}
\usepackage{xcolor}
\usepackage{microtype}
\usepackage{subcaption}
\usepackage{balance}
\usepackage{mathtools}
\usepackage{scalerel}
\usepackage{adjustbox}
\usepackage{float}

\usepackage[noend]{algpseudocode}
\usepackage{algorithmicx,algorithm}

\definecolor{darkgreen}{rgb}{0.15,0.55,0.15}
\definecolor{darkblue}{rgb}{0.1,0.1,0.5}
\definecolor{orange}{rgb}{1, 0.64, 0}
\definecolor{blue}{rgb}{0.01,0.40,.8}
\definecolor{darkgreen}{rgb}{0.15,0.55,0.15}
\definecolor{mred}{rgb}{.80,.12,.30}
\definecolor{grey}{rgb}{0.5,0.5,0.5}
\definecolor{Purple}{rgb}{.75,0,.85}
\definecolor{light-gray}{gray}{0.95}
\definecolor{mid-gray}{gray}{0.85}
\definecolor{darkred}{rgb}{0.7,0.25,0.25}
\definecolor{rose}{rgb}{1.0, 0.01, 0.24}
\definecolor{lightyellow}{HTML}{FFD700}
\definecolor{lgreen}{HTML}{C2E9D7}

\newcommand{\gray}[1]{\textcolor{grey}{#1}}

\newcommand{\highlight}[1]{{\setlength{\fboxsep}{1pt}\colorbox{yellow}{\textbf{\texttt{#1}}}}}

\newcommand{\eat}[1]{}

\newlength{\listingindent}                %
\usepackage{tablefootnote}

\usepackage[LGRgreek]{mathastext}
\newtheorem{example}{Example}

\definecolor{applegreen}{rgb}{0.55, 0.71, 0.0}
\definecolor{asparagus}{rgb}{0.53, 0.75, 0.48}
\definecolor{deeplilac}{rgb}{0.6, 0.33, 0.73}
\definecolor{maroon}{rgb}{0.5, 0.0, 0.0}
\definecolor{royalfuchsia}{rgb}{0.79, 0.25, 0.37}
\definecolor{maroon}{rgb}{0.69, 0.19, 0.38}

\newcommand{\sysold}{\textsc{PI2}\xspace}

\newcommand{\sys}{\textsc{NL2Interface}\xspace}
\newcommand{\difftree}{\textsc{Difftree}\xspace}
\newcommand{\difftrees}{{\difftree}s\xspace}
\newcommand{\sps}{\textsc{SPS}\xspace}

\title{\sys: Interactive Visualization Interface Generation from Natural Language Queries}

\author{Yiru Chen\thanks{e-mail: yiru.chen@columbia.edu}\\ %
        \scriptsize Columbia University %
\and Ryan Li\thanks{e-mail: lansong@cs.washington.edu}\\ %
     \scriptsize University of Washington 
 \and Austin Mac \thanks{e-mail: austinmac@ucsb.edu}\\ %
     \scriptsize University of California, Santa Barbara
      \and Tianbao Xie\thanks{e-mail: tianbaox@cs.hku.hk}\\ 
 \scriptsize University of Hong Kong
 \and Tao Yu\thanks{e-mail: tyu@cs.hku.hk}\\ 
 \scriptsize University of Hong Kong
 \and Eugene Wu\thanks{e-mail: ewu@cs.columbia.edu}\\ %
     \scriptsize Columbia University
     }

\teaser{
\vspace{2mm}
\center
 \includegraphics[width=\textwidth]{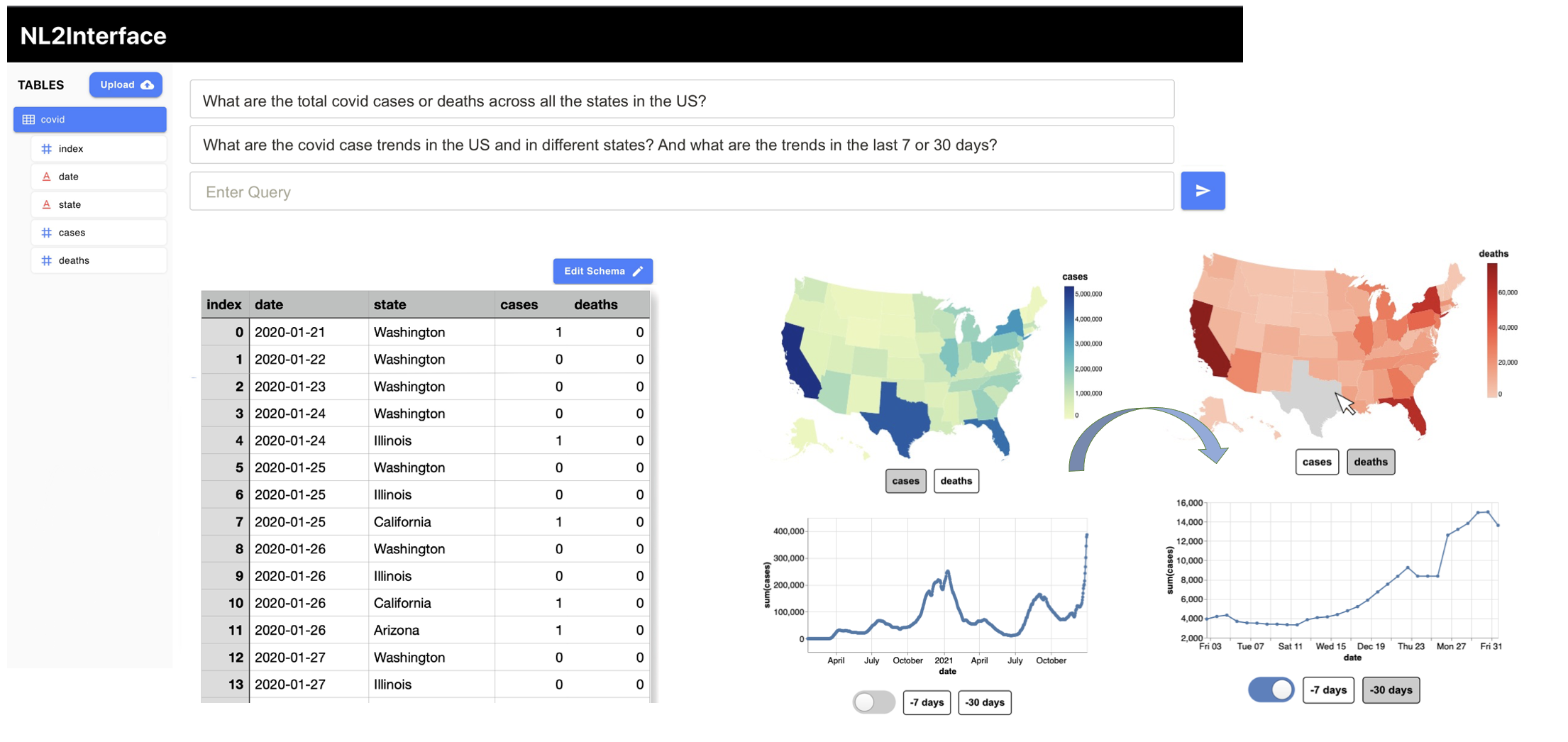}
 \caption{This demonstrates the \sys user interface. A user uploads the COVID-19 dataset using the left panel and the table contents are displayed in the bottom left of the interface. Then, this user types two natural language queries in the top textbox. Our \sys will return an interactive interface consisting of two visualizations, one toggle, two button sets, and a click interaction over the map visualization.  The user can interact with the interface to study the covid death trend within the last 30 days in Texas by clicking the state of ``Texas” on the map, clicking the ``deaths” button, and then clicking the ``-30 days” button. The interface after these interactions is shown on the right. } 

 \label{example}

}

\abstract{
  We develop \sys to explore the potential of generating usable interactive multi-visualization interfaces from natural language queries.  With \sys, users can directly write natural language queries to automatically generate a fully interactive multi-visualization interface without any extra effort of learning a tool or programming language. Furthermore, users can interact with the interfaces to easily transform data and quickly see results in the visualizations.  
} %

\begin{document}

\maketitle

\section{Introduction}

Interactive visualization interfaces (or simply {\it interfaces}) play a critical role in nearly every stage of data management––including data cleaning~\cite{wu2013scorpion}, wrangling~\cite{Kandel2011WranglerIV}, modeling~\cite{facets}, exploration~\cite{Murray2013TableauYD, Chen2021TSExplainSE}, and communication~\cite{icheck, fivethirtyeight}.
A specific data management task is a series of data transformations commonly performed with SQL that gradually converts data into an ideal format.  
The interface design primarily focuses on visualizations and providing interactions. Interactions including widgets and visualization interactions such as brushing and clicking help users easily specify the desired transformations and let users quickly see results in the visualizations. 

Such interfaces require considerable expertise and trial-and-error to design and implement. This is because end users have to determine the data management task, then choose visualizations, interactions, and layout to express the underlying data transformations~\cite{munzner2014visualization}.

Existing works ~\cite{cox2001multi, sun2010articulate, aurisano2016articulate2, gao2015datatone, hoque2017applying, narechania2020nl4dv} allow users to express data transformations in natural language queries and return data visualizations. Using natural language queries for data transformation helps users express the transformation directly without having to learn a tool or a programming language. This is desirable for normal users, especially non-programmers. 
However, existing works can only create individual visualizations instead of interactive multi-visualization interfaces. 
Unlike an individual visualization, creating an interactive multi-visualization interface requires much more consideration on  widgets choices, layout design, number of visualizations, whether or not to allow interactions on visualizations, types of interactions, etc. 

All the above considerations make it very hard to generate interfaces from natural language queries. 
One key challenge includes: {\it  What is the target representation of the data transformations for natural language query input? } One natural language query may specify multiple data transformations. For example, a natural language query - ``What is the covid trend in different states?" includes multiple different queries varying the state filter. Thus, a new representation is needed to represent the underlying data transformations of natural language queries.
After designing the representation,   we still need to {\it translate natural language queries into such a representation, and generate interfaces from such a representation}.

In this paper, we develop {\bf \sys  to explore the potential of generating usable interactive multi-visualization interfaces from natural language 
queries}.
Under the hood, we use SQL as the data transformation language.  Adapted from a recent representation in \sysold, a SQL-based interface generation system~\cite{pi2demo, chen2022pi2}, we extend SQL with a parameterized syntax to encode the structural similarities and differences between SQL queries. We name it structurally parameterized SQL (\sps).
With advancements in large language models~\cite{scholak2021picard, https://doi.org/10.48550/arxiv.2109.05153, shaw2020compositional,chen2021evaluating, brown2020language}, we use  a transformer pre-trained language model named Codex~\cite{chen2021evaluating} to translate natural language to \sps. Codex only needs a small number of natural language queries and \sps examples as prompts to predict the \sps representation.
Afterwards, we use \sysold to automatically generate fully interactive interfaces from the codex-predicted \sps.

Below is an end-to-end example: 

\begin{example}[Covid Analysis]\label{covid}
\Cref{example} shows the \sys user interface. The left panel allows users to upload a dataset and shows the table schemas. The table contents are shown to the right.  Users can type one or multiple natural language queries using the textbox. After clicking the send button, \sys returns an interface for users to interact with. 

Here, we walk through how to use \sys to generate a useful interface using a COVID-19 dataset. 
First,  the user uploads the COVID-19 dataset using the upper left {\tt upload} button. Then, they can view the table schema directly below the {\tt upload} button and the table contents in the bottom left of the UI. 
Given this dataset, the user is interested in two natural language queries: ``What are the total covid cases or deaths across all the states in the US?", and ``What are the covid case trends in the US and in different states? And what are the trends in the last 7 or 30 days?" The user types these queries in the text box at the top of the \sys interface and clicks the send button. 

\sys generates an interactive visualization interface consisting of two visualizations, one toggle, two button sets, and a  click interaction in the top map visualization. The map visualization corresponds to the first query. Users can interact with the button widgets in the middle of the interface to choose between cases and deaths. The line chart below shows the overall trend in the US. Users can toggle on to specify the date range (e.g. last 30 days), or users can toggle off to see the trend for all data points. 
Besides widgets, users can click on the map visualization to filter specific states (e.g. ``Texas''). The interface after user interactions is shown on the right where the line chart updates to show covid cases for the last 30 days in Texas.

\end{example}

As shown above, \sys lets users directly input their natural language queries, and automatically generates a  fully interactive multi-visualization interface without any extra effort of learning a tool or programming language.

\section{Related Work}
\label{related}

\subsection{Natural Language Interface for Data Visualization}
Natural language interfaces for data visualization have been studied extensively~\cite{cox2001multi, sun2010articulate, aurisano2016articulate2, gao2015datatone, hoque2017applying, narechania2020nl4dv}. Unlike \sys which uses natural language dialogues to create multi-visualization interfaces, existing works are focused on using natural language dialogues to create individual visualizations. 

Cox et al.~\cite{cox2001multi} is an early work which takes a natural language question, determines the appropriate database query, and presents a visualization to the user.
The Articulate system~\cite{sun2010articulate} and Articulate2~\cite{aurisano2016articulate2}  use a conversational user interface to allow users to verbally describe what they want to see. This description is then translated into sketch, analysis, and manipulation commands.  
Similarly, DataTone~\cite{gao2015datatone} generates an individual visualization for natural language queries. Furthermore, it can manage the ambiguity in natural language input as widgets.
Eviza~\cite{setlur2016eviza} and Evizeon~\cite{hoque2017applying} investigate natural language as a means of interacting with existing visualizations. NL4DV~\cite{narechania2020nl4dv} is a toolkit that supports prototyping natural language interfaces for visualizations. Given an NL query, NL4DV generates visualizations by parsing the NL query to extract attributes and tasks.

In contrast, \sys creates interactive multi-visualization interfaces which consider visualizations, interactions and layout. None of the existing works have explored this.

\subsection{Natural Language to SQL Translation}
Translation from natural language to SQL (Text-to-SQL) has been widely studied by the NLP community~\cite{Yu18c, wang2020rat, shaw2020compositional, scholak2021picard, rajkumar2022evaluating, qi2022rasat}. 
Difficulties in text-to-SQL are mainly two-fold: encoding a variety of complex relationships between the user's query and multiple tables, and decoding the SQL with valid representations.
Many successful solutions include question-table encoding and constrained decoding.

Recently, Shaw et al.~\cite{shaw2020compositional} showed that fine-tuning a pre-trained T5~\cite{Raffel2020ExploringTL} language model could yield competitive results on Spider~\cite{yu2018spider} text-to-SQL benchmark.
PICARD~\cite{scholak2021picard} constrains the auto-regressive decoder during inference time to improve the quality of SQL generation.
More recently, large language models (LLMs) like GPT-3~\cite{brown2020language} and Codex~\cite{chen2021evaluating} have been shown to perform incredibly well in many NLP tasks without any training.
\cite{rajkumar2022evaluating} demonstrates Codex's near state-of-the-art performance on Spider in a zero-shot setting when prompted with in-context examples. 

In this work, \sys uses Codex as the back-end LLM. 
\cite{Shin2022FewShotSP} finds that Codex is able to generalize to unseen target programs with a few in-context demonstration examples.
In our case, we prompt Codex to generate our designed SPS representations given natural language inputs by providing a few natural language-SPS pairs.

\subsection{Precision Interface}

The Precision Interface series of work~\cite{Zhang2018PrecisionIF, Chen2020MonteCT, chen2022pi2} explores interface generation from queries. 
PI~\cite{zhang2019mining} uses SQL queries as a proxy and generates interactive widgets from a sequence of input queries. At a high level, it identifies structural differences between queries and maps them to widgets. However, it does not consider visualizations, visualization interactions, nor layout.
\sysold~\cite{chen2022pi2,pi2demo} automatically generates fully interactive multi-visualization interfaces from SQL analysis. \sysold proposes \difftrees as a compact representation encoding the structural similarities and  differences between SQL abstract syntax trees and searches for a good interface for the SQL analysis given a cost model. Our \sps representation design is based on \difftrees; however, the difference is that  \sps is SQL-like text strings rather than a tree structure.

Compared with \sysold, \sys is able to generate interfaces from natural language queries which makes it more accessible to everyone, especially non-programmers. 
\section{\sys System}

\begin{figure}[H]
    \centering
    \includegraphics[width=\columnwidth]{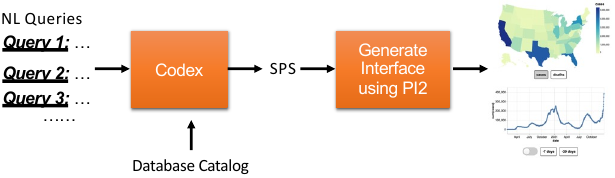}
    \caption{The pipeline for generating interactive interfaces from natural language queries.}

    \label{fig:pipline}
\end{figure}

\sys is built upon OpenAI's Codex~\cite{chen2021evaluating} and our previous work \sysold~\cite{chen2022pi2} – the interface generation system from SQL analyses.  \Cref{fig:pipline} shows the overall pipeline. First, \sys prepares a few examples of translating from natural language queries to structurally parameterized SQL (\sps) representations, to build a proper prompt for in-context learning for Codex. Then, given natural language queries and a database catalog, Codex predicts \sps representations of the natural language queries. Finally, \sys follows the interface mapping procedure in \sysold~\cite{chen2022pi2} to generate interfaces based on a predefined yet extensible cost model.

Below, we will introduce the \sps representation, how to use the Codex model to predict \sps, and the interface generation for \sps using \Cref{covid}.

\subsection{Structurally Parameterized SQL (\sps)}

Structurally Parameterized SQL (\sps) is a SQL-like text which extends SQL with choice nodes - {\tt ANY}(choose one of many), {\tt OPT}(exist or not), and {\tt SUBSET}(choose a subset from all choices). It can efficiently represent a large set of queries by extracting and encoding the variations between the queries. 
\sps's choice nodes can be used to choose from literal parameters, expressions, SQL clauses, or even subqueries. Though \sps is similar to parameterized SQLs~\cite{paraSQL}, the expressiveness of \sps expands from supporting literal parameters as parameterized SQLs~\cite{paraSQL} to any query expressions, clauses, or subqueries. 

Here, we define the choice node syntax:

\begin{enumerate}
\item \texttt{ANY\{c1,..,ck\}} can choose one of its $k$ choices between the parentheses, akin to the production rule \texttt{ANY$\to$c1|..|ck}. 
Additionally, to specify a range of numeric values, one can use {\tt ANY\{num1-num2\}}; e.g., {\tt ANY\{3.0-5.0\}} represents any numeric value between 3.0 and 5.0. One can also use {\tt \&attr} to refer to all distinct values of this attribute in the database. For example, {\tt ANY\{\&state\}} indicates that this {\tt ANY} can choose between all different states in the data.

\item \texttt{SUBSET[sep]\{c1,..,ck\}} represents the production rule \texttt{SUBSET$\to$c1?..ck?} with \texttt{sep} as the separator.   {\tt SUBSET} can choose a subset of its choices listed between the parentheses connected by {\tt sep}. 
The attribute domain shortcut {\tt \&attr} is also applicable for {\tt SUBSET}.  
For example, {\tt state in (SUBSET[,]\{\&state\})} allows users to choose any set of states in the attribute {\tt state}'s domain.

\item \texttt{OPT\{t\}$\to$t|NULL}. The OPT choice node implies that {\tt t} is optional in the query. This can be viewed as a special case of {\tt ANY} choice node where one choice is empty. 
\end{enumerate}

\begin{lstlisting}[caption = { SQL queries and their corresponding \sps for the first natural language query in \Cref{covid}.}, label={l:covid}]
    q1: select state, ^\highlight{sum(cases)}^ from covid group by state  
    q2: select state, ^\highlight{sum(deaths)}^ from covid group by state 
    SPS: select state, sum(^\highlight{ANY\{cases, deaths\}}^) from covid group by state 
\end{lstlisting}

\begin{example}[\sps]
The first natural language query in \Cref{covid} -``What are the total covid cases or deaths across all the states in the US?"  contains two queries - {\tt q1} and {\tt q2} shown in \Cref{l:covid}. The \sps representation shown in \Cref{l:covid} uses an {\tt ANY} choice node to let users choose between cases and deaths. 
\end{example}

\subsection{Codex: Natural Language Queries to \sps}  
\sys uses Codex to translate natural language queries to \sps representations. Though natural language queries could be arbitrary~\cite{zhang2020did} and even irrelevant to the dataset, we currently focus on natural language queries that correspond to data transformations on the given dataset. We leave detecting inappropriately structured queries and handling unanswerable questions as future work.  

Codex is a GPT language model trained on both natural language and publicly available GitHub code \cite{chen2021evaluating}. 
We choose Codex because, given a few examples and without any finetuning, Codex has proven to be comparable to state-of-the-art fine-tuned systems trained on thousands of annotated text-SQL pairs~\cite{rajkumar2022evaluating}, which fits our needs best, since we do not have large-scale training data to re-train a text-to-SPS model.

More specifically, Codex has several model series with different parameter sizes, and we use the \textit{davinci-code-002} model, which is the largest and the most powerful model instance of Codex.
Few-shot learning with Codex leverages a recent learning paradigm, in-context learning, and doesn't require any training or fine-tuning.
Codex observes a test natural language query and a few text-to-SPS input-output demonstration examples as its input, and directly decodes the corresponding SPS program without any update to its parameters.
To format the text-to-SPS input-output demonstration inputs, we follow~\cite{rajkumar2022evaluating} and provide 5 database schema, each representing a distinct domain: weather, flights, sales, cities, and cars. 
Under each database schema, we include 10 different natural language to \sps translation examples for that database. 
The examples are designed to cover the basic usage of {\tt ANY}, {\tt SUBSET}, and {\tt OPT} choice nodes to provide Codex a solid overview of NL to \sps translation. 
Each example is an  {\tt <NL query, \sps>} pair.

Once the user uploads a database and sends a natural language query, we concatenate the schema of the database and the natural language query as the test instance and then append it to the above text-to-SPS input-output demonstration inputs to create a final inference input prompt to Codex.
Then the Codex \textit{davinci-code-002} model generates an \sps for each NL query based on the prompt provided. 
\begin{lstlisting}[caption = {
The second natural language query and its corresponding \sps in \Cref{covid}.}, label={l:covid2}]
NL Query: What are the covid case trends in the ^\highlight{US}^ and in ^\highlight{different states}^? And what are the trends in the last ^\highlight{in 7 or 30 days}^?

SPS: 
  select date, sum(cases) from covid 
  where ^\highlight{OPT\{state = ANY\{\&state\}\}}^ and  
  ^\highlight{OPT\{date > date(today(), ANY\{'-7 days', '-30 days'\})\}}^
  group by date

\end{lstlisting}
\begin{example}[Codex Prediction]
\Cref*{l:covid2} shows the translated \sps for the second NL query in \Cref{covid}. In the \sps, {\tt OPT\{state = ANY\{\&state\}\}} corresponds to different states' trend. If {\tt OPT} is chosen, it will show a certain state's trend specified by the {\tt ANY} choice node. {\tt \&state} represents all distinct values of attribute {\tt state} in the database. Otherwise, it will show the US overall trend. 
 {\tt OPT \{date > date(today(), ANY\{`-7 days', `-30 days'\})\}} toggles the date range filter. When it is on, the {\tt ANY} node will specify the last 7 days or 30 days. Otherwise, it will show the trend for all time.
 \end{example}
 
In our preliminary work, we evaluate Codex performance by choosing three different databases unseen by Codex and designing four natural language queries for each database. 
We found that the \textit{davinci-code-002} is able to correctly translate 8 out of the 12 natural language queries into accurate, executable \sps. 
Further improvements to the prompt and a systematic evaluation of Codex's performance will be conducted in our future work.

\subsection{Interface Generation}
After translating natural language queries to Structurally Parameterized SQLs (\sps), we follow the interface mapping procedure in \sysold~\cite{chen2022pi2} to generate the interface from these SPSs. \sys chooses a visualization for each \sps, chooses widgets or adds visualization interactions to parameterize the choice nodes, and chooses a layout that takes the screen size into account. The visualizations are chosen using a heuristic based on prior work~\cite{Mackinlay2007ShowMA} but can use recent visualization recommendations as well. PI2 uses a randomized search procedure and uses a simple yet extensible cost model that takes user effort into account~\cite{Gajos2004SUPPLEAG}.

\begin{example}[Interface Generation]
In \Cref{covid}, the first \sps project attribute shown in \Cref*{l:covid} is {\tt state, sum(ANY\{cases, deaths\})}. The map visualization renders the query result as a choropleth, while a button below chooses the statistic-total {\tt cases} or {\tt deaths} to encode as each state's color as shown in \Cref{example}. 
The \sps in \Cref{l:covid2}'s project attributes are {\tt date, sum(cases)}. \sys maps this \difftree to a line chart in \Cref{example}. Interestingly, given the first Map visualization,  {\tt OPT\{state = ANY\{\&state\}\}} can be directly expressed by the click interaction on the map. Once clicked on the above map, the {\tt OPT} node is turned on and the clicked state will parameterize the {\tt ANY} choice node. For example, users can click on the ``Texas" state on the map and the line chart will show the Texas trend.  If the map is not clicked, the {\tt OPT} node is turned off and the line chart will show the overall trend in the US. Such a visualization interaction (clicking over the map) has multiple benefits: it makes the interface more concise; it lets users directly interact with the map visualization to specify the transformation; it potentially builds a relation between two visualizations to enable more insights, etc. Finally, \sys maps the last part, {\tt OPT\{date > date(today(), ANY\{'-7 days', '-30 days'\})\}}, to a toggle corresponding to {\tt OPT} and a button corresponding to {\tt ANY}.  
Now, \sys can return the generated interface to users.
\end{example}

\section{Discussion}

\sys is the first to prove the feasibility of automatically generating interfaces from natural language queries. It takes in user specified natural language queries, translates the queries into \sps representations using Codex, and then maps the \sps representations to an interactive multi-visualization interface. 
However, there are still some limitations.

First, the natural language queries \sys accepts are restricted to those corresponding to data transformation on the given database. However, the NL queries could be arbitrary, irrelevant, or ambiguous~\cite{zhang2020did}.
Given recent progress in large pretrained language models, where a single language model is able to achieve state-of-the-art few-shot performance on a wide range of NLP tasks, it is possible to design a pipeline that handles all possible kinds of user queries and reacts correspondingly.
For example, Codex can perform a task that detects which queries cannot be mapped to SQL queries given a database and asks the user to ask questions answerable by the database.
Besides data transformations, it is also interesting to express the interface preferences in the NL queries, e.g. which visualization to use for a certain NL query to improve the generated interface. 

Second, we heavily rely on the prompt design, although Codex is already a powerful model in generalization cross task and domain.
Carefully designing the prompt format and examples for forming the prompt is still essential to the performance of prediction~\cite{min2022rethinking}.
For now, we manually choose examples, make them as diverse as possible at the beginning, and alter them incrementally according to their performance in development in order to maximize performance.
The research of automatically selecting proper examples and organizing them a certain way to maximize the ability of Codex is another important direction.

Also, \sys can be used as an interactive human-centered method to allow laypeople to verify the predictions of  text-to-SQL neural models such as Codex by checking the resulting visualization. 
Despite the surprising performance of recent deep learning text-to-SQL methods, interoperability is still one of the key concerns people have with these models. 
Mapping the predicted target programs to visual interfaces provides users without a technical background a way to verify if these text-to-SQL neural models understand their questions correctly.

\acknowledgments{
Thanks to Alexander Yao, Jeffrey Tao, Brandon Zhang, and Xiaoyue Chen for early contributions. This work was partially supported by NSF 1845638, 1740305, 2008295, 2106197, 2103794, Amazon, Google, and Columbia SIRS.
}

\bibliographystyle{abbrv-doi}

\bibliography{ref}
\end{document}